\documentclass{llncs}

\usepackage{graphicx}
\usepackage{epsfig}
\usepackage{amsmath}
\usepackage{amssymb}
\usepackage{cite}

\newsavebox{\SquareEmpty}

\newsavebox{\SquareWithX}

\newsavebox{\SquareFull}

\newsavebox{\SquareBarsA}

\newsavebox{\SquareBarsB}

\newtheorem{fact}{Fact}

\begin{document}

\pagestyle{headings}

\mainmatter

\title{On Short Cuts \\ or \\ Fencing in Rectangular Strips}

\titlerunning{On Short Cuts or Fencing in Rectangular Strips}

\author{Yaniv Altshuler\inst{1}
\and Alfred M. Bruckstein\inst{2}}

\authorrunning{Y.Altshuler, V.Yanovsky and A.M.Bruckstein}

\institute{Deutsche Telekom Laboratories and Information Systems Engineering Department, Ben Gurion University, Beer Sheva 84105, Israel \\
\email{yanival@cs.technion.ac.il}
\and Computer Science Department, Technion, Haifa 32000 Israel \\ %
\email{freddy@cs.technion.ac.il}
}

\maketitle

\begin{abstract}
In this paper we consider an isoperimetric inequality for the \emph{free perimeter} of a planar shape inside a rectangular domain, the free perimeter being the length of the shape boundary that does not touch the border of the domain.
\end{abstract}

\section{Introduction}

The isoperimetric inequality for shapes in $\mathbb{R}^{2}$ states that the area
enclosed by a simple closed curve is at most that of a
circle of the same length, and that equality occurs only for circles. This immediately implies that among all simple closed curves enclosing a given area, a circle is the shortest.

Several variations on the isoperimetric inequality were considered in the literature (see e.g. \cite{RO1979}). In this paper we shall discuss inequalities involving the notion of ``\emph{free perimeter}'' for a shape $S$, located inside a simple, bounded, planar domain $D$. We may assume that there is a border or wall surrounding this domain, or alternatively that this domain is an island surrounded by water.
A simple shape $S$ inside this domain will be defined by a boundary curve, some portion of which may touch and even follow the border (wall / shoreline) of the domain / island. The free perimeter of the shape will be defined as the length of the boundary curve of $S$ that does not overlap with, or trace, the border of the enclosing domain $D$.

The problem that we can pose with these definitions is the following : given the domain $D$, determine the shape with the shortest free-perimeter that has a given area $A$. This problem is, of course, that of determining the way to cut out a shape of a total area $A$ from $D$ with the least effort of cutting, i.e. with the \emph{shortest cut}. Equivalently, this is the problem of determining the shortest length ``fence'' that can separate a contiguous region of area $A$ inside the domain $D$.

This interpretation clearly explains the totally misleading title of our paper, in which we do not take any short-cuts and of course we do not discuss fencing as a sport that happens to be played on a rectangular, strip-shaped, ``ring''.

A related problem is that of finding the connected shape of largest area that can be ``lifted out'' of $D$ with a total length of ``cuts'' or ``fences'' less than or equal to $L$.

In this paper we solve the problem raised above when the region $D$ is a rectangle. We prove that the shortest cut, i.e. the minimum free perimeter, that separates a shape with half of the area of $D$ has, as expected, the length of the shorter side of the rectangle. We then provide the shortest free perimeter for all $\frac{Area(S)}{Area(D)}$ ratios from 0 to 1.

We note that the problem we discuss is closely related to the problem A26, ``\emph{Dividing up a piece of land by a short fence}'', discussed in the book ``\emph{Unsolved Problems in Geometry}'' \cite{CFG1991}. The challenge posed there is that of dividing a convex shape into two equal-area parts. We refer the interested reader to \cite{CFG1991} and to some recent follow-up papers \cite{MPS2004,GKMS2007}.

\begin{figure}[htb]
   \centering
   \includegraphics[clip=true, trim=0in 0in 0in 1.5in,scale=0.35,bb=0 0 800 500]{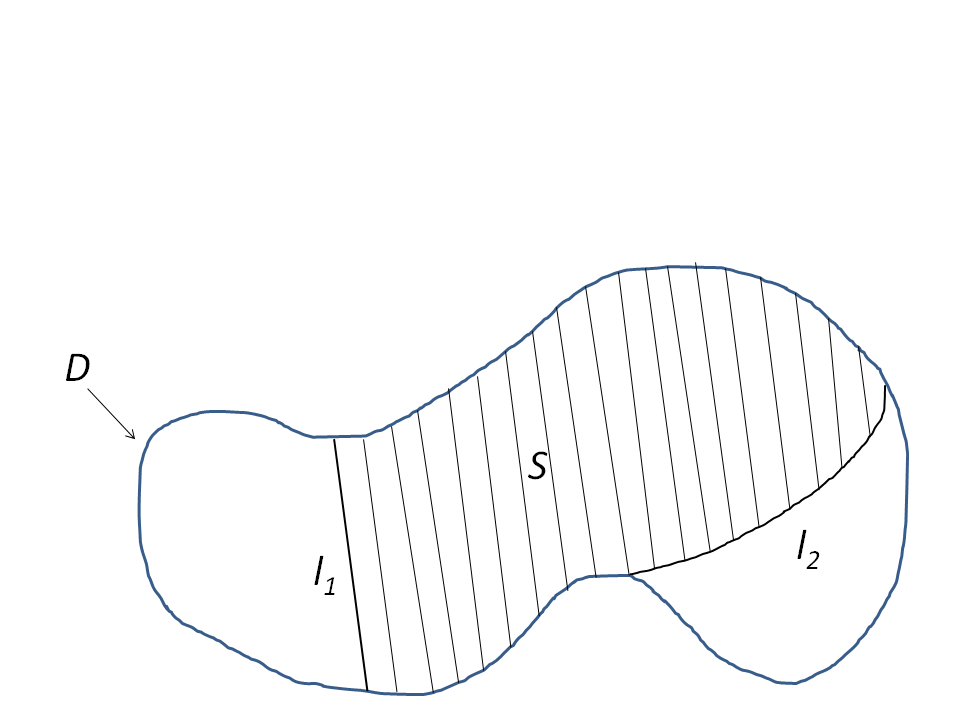}
   \caption{An illustration of the notion of ``\emph{free perimeter}''. The area of the shape $S$ equals A, while its free perimeter equals $l_{1} + l_{2}$.
   }
   \label{fig1}
\end{figure}

\section{Free Perimeter of Half Area Shapes in Rectangles}\label{sec.half}

Let $D[X,Y]$ be a bounding rectangle of dimensions $X$ and $Y$, with $X \leq Y$. Let $A$ be the area we want to enclose with a region of shape $S$, and denote by $l_{FP}(S)$ the length of the free perimeter of the shape $S$. Let us denote by $l^{\star}(A)$ the length of the free perimeter of a shape $S$ with area $A$, such that $S$ has the smallest value of $l_{FP}(S)$ out of all the shapes of area $A$. Namely~:

\[
l^{\star}(A) \triangleq \min_{Area(S) = A} \{ l_{FP}(S) \}
\]

We shall be interested in determining the value of $l^{\star}(A)$ for $A \in [0,XY]$.
For this, we shall first prove the following result~:

\begin{theorem}  \label{thm1}
\[l^{\star}\left(A = \frac{1}{2} XY\right) = X\]
\begin{proof}
To prove the above stated, and rather natural and hardly surprising result, we shall need to combine several simple facts.

\begin{fact} \label{lemma1} \textbf{\emph{The Classical Planar Isoperimetric Inequality}} \\
Given any shape of area $A$ in the plane, and perimeter of length $l$ we have~:
\[
l \geq 2 \sqrt{\pi} \sqrt{A} = \sqrt{4 \pi A}
\]
with equality achieved for a circle.
\end{fact}

\begin{fact} \label{lemma2} \textbf{\emph{The Half-Plane Isoperimetric Inequality}} \\
Given any shape $S$ of area $A$ in a half plain domain, with free perimeter of $l_{FP}(S)$ we have~:
\[
l_{FP} \geq \sqrt{2 \pi A}
\]
\begin{proof}
If $S$ touches the boundary of the half-plane, let us reflect it along the boundary line, thereby generating a (symmetric) shape of area $2A$ in the plane. For this ``double shape'' $S'$ we have~:
\[l_{FP}(S') = 2 l_{FP}(S)\]
and with the classical isoperimetric inequality of Fact \ref{lemma1} we obtain~:
\[l_{FP}(S') \geq 2 \sqrt{\pi} \sqrt{2A}\]
hence~:
\[l_{FP}(S) = \frac{1}{2} l_{FP}(S')\geq \sqrt{\pi} \sqrt{2A}\]
\end{proof}
\end{fact}

\begin{fact} \label{lemma3} \textbf{\emph{The Quarter-Plane Isoperimetric Inequality}} \\
Given any shape $S$ of area $A$ in a quarter plain domain, with free perimeter of $l_{FP}(S)$ we have~:
\[
l_{FP} \geq \sqrt{\pi A}
\]
\begin{proof}
If $S$ touches the two orthogonal boundaries of the quarter-plane, let us reflect it symmetrically into the three quarters plane domain boundary, generating a shape $S'$ in the plane, of area $4A$. For $S'$ we have~:
\[l_{FP}(S') = 4 l_{FP}(S)\]
and with the classical isoperimetric inequality of Fact \ref{lemma1} we obtain~:
\[l_{FP}(S') \geq 2 \sqrt{\pi} \sqrt{4A}\]
yielding~:
\[l_{FP}(S) = \frac{1}{4} l_{FP}(S')\geq \sqrt{\pi} \sqrt{A}\]
\end{proof}
\end{fact}

A shape $S \subset D[X,Y]$ may touch the sides of the boundary of the rectangle $D(X,Y)$ in several ways. We may have $S$ that touches 0,1,2,3 or 4 sides. Let us consider these cases separately~:

\noindent \textbf{Case 0 :} $S$ touches 0 sides of $D[X,Y]$. In this case, the classical isoperimetric inequality of Fact \ref{lemma1} yields~:
\[l_{FP}(S) \geq 2 \sqrt{\pi} \sqrt{\frac{1}{2} XY} \geq \sqrt{2\pi} \sqrt{XY} \geq \sqrt{2\pi} \cdot X > X\]

\noindent \textbf{Case 1 :} $S$ touches 1 of the sides of $D[X,Y]$. In this case, Fact \ref{lemma2} yields~:
\[l_{FP}(S) \geq \sqrt{2 \pi} \sqrt{\frac{1}{2} XY} \geq \sqrt{\pi} \sqrt{XY} \geq \sqrt{\pi} \cdot X > X\]

\noindent \textbf{Case 2 :} $S$ touches 2 of the sides of $D[X,Y]$. In this case we have either $S$ touches two opposite sides, yielding $l_{FP}(S) \geq  2 \min \{X,Y\} \geq 2X$, or $S$ touches two adjacent sides, in which case Fact \ref{lemma3} provides~:
\[l_{FP}(S) \geq \sqrt{\pi} \sqrt{\frac{1}{2} XY} \geq \sqrt{\frac{1}{2}\pi} \sqrt{XY} \geq \sqrt{\frac{1}{2}\pi} \cdot X > X \qquad \emph{(since $Y \geq X$)}\]

\noindent \textbf{Case 3 :} $S$ touches 3 of the sides of $D[X,Y]$. In this case we have $l_{FP}(S) \geq \min \{X,Y\} \geq X$, since any of the portions of the boundary of $S$ will have to join parts on opposite sides of $D[X,Y]$.

\noindent \textbf{Case 4 :} $S$ touches all four sides of $D[X,Y]$. In this case we have a connected shape $S$ which is continuous (i.e. connected), whose complement $S^{C} \triangleq D[X,Y] \setminus S$ might be a set of disconnected regions $S^{C}_{1}, S^{C}_{2}, S^{C}_{3}, \ldots, S^{C}_{k}$, of areas $A_{1}, A_{2}, A_{3}, \ldots, A_{k}$, which all belong to $D[X,Y]$, and for which we have~:
\[
\sum A_{i} = \frac{1}{2} XY
\]

We also have that~:
\[
\sum l_{FP}(S^{C}_{i}) = l_{FP}(S^{C}) \equiv l_{FP}(S)
\]

Notice that for all $i$, $S^{C}_{i}$ cannot touch more than 2 sides of the rectangle $D[X,Y]$, since this would imply that $S$ is disconnected.

By Facts 1,2 and 3 we therefore have~:
\[
f_{FP}(S^{C}_{i}) \geq \min\{\sqrt{\pi},\sqrt{2 \pi},\sqrt{4 \pi}\} \cdot \sqrt{A_{i}} = \sqrt{\pi} \sqrt{A_{i}}
\]
and subsequently~:
\[
f_{FP}(S) = f_{FP}(S^{C}) = \sum_{i = 1}^{k} l_{FP}(S^{C}_{i}) \geq \sqrt{\pi} \sum_{i = 1}^{k} \sqrt{A_{i}}
\]

Notice that~:
\[
\left( \sum_{i = 1}^{k} \sqrt{A_{i}} \right)^{2} = \underbrace{\sum_{i = 1}^{k} A_{i}}_{=A} + \sum_{i \neq j} \sqrt{A_{i}} \sqrt{A_{j}}
\]

Hence~:
\[
\sum_{i = 1}^{k} \sqrt{A_{i}} \geq \sqrt{A}
\]
and therefore~:
\[
l_{FP}(S) \geq \sqrt{\pi} \sum_{i = 1}^{k} \sqrt{A_{i}}  \geq \sqrt{\pi} \sqrt{A} \geq \sqrt{\pi} \sqrt{\frac{1}{2} XY} \geq X
\]

It is important to note that although~:
\[
l_{FP}(S) \geq \sqrt{\pi} \sum_{i = 1}^{k} \sqrt{A_{i}}  \geq \sqrt{\pi} \sqrt{A}
\]
in fact~:
\[
l_{FP}(S) \geq \sqrt{\pi} \sum_{i = 1}^{k} \sqrt{A_{i}}  \geq \sqrt{\pi} \sqrt{Area(D) - A}
\]
(which is the same in this case, as here $Area(D) = 2A$).

We have shown that in all cases, $l_{FP}(S) \geq X$. It is easy to see that when $S$ is defined as the half-rectangle $X \times \frac{1}{2}Y$, the free perimeter obtained is exactly $X$. Therefore, we have shown that $l^{\star}(\frac{1}{2}XY) = X$.
\end{proof}
\end{theorem}

\begin{figure}[htb]
   \centering
   \includegraphics[scale=0.25,bb=0 0 900 650]{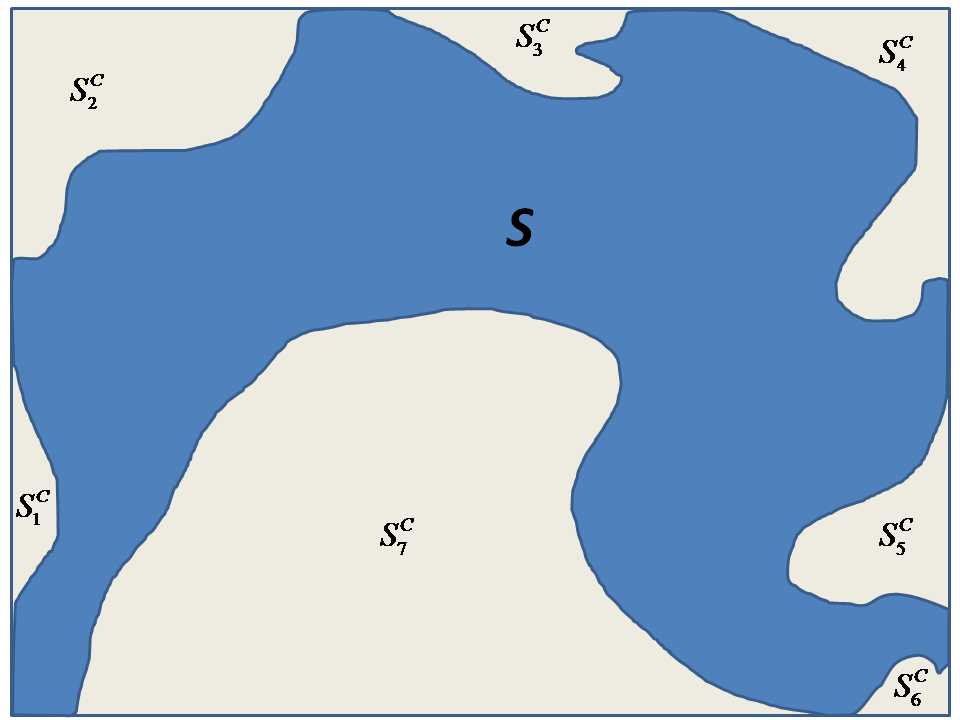}
   \caption{An illustration of a shape $S$ that touches all four sides of the rectangle $D[X,Y]$ and its complement $S^{C}$ that comprised out of a set of connected regions $S^{C}_{i}$.
   }
   \label{fig1}
\end{figure}

In fact, we have shown something stronger that just $l^{\star}(\frac{1}{2}XY)=X$. In all cases where $S$ touches 0, 1, 2 or 4 sides of the rectangle, its free perimeter $l_{FP}(S)$ was strictly higher than $X$, by factors of $\sqrt{2\pi} > 2 > \sqrt{\pi} > \sqrt{\frac{\pi}{2}} > 1$.

Interestingly, note that $\sum_{i=1}^{k} \sqrt{A_{i}}$ is maximized where $\forall i, A_{i} = \frac{A}{k}$~:
\begin{proof}
Let us define ~:
\[\Psi = \sum_{i=1}^{k} \sqrt{A_{i}} + \lambda \left( \sum_{i=1}^{k} A_{i} - A\right)\]

In order for $\frac{\partial \Psi}{\partial A_{i}} = 0$ we must have $\frac{1}{2} \frac{1}{\sqrt{A_{i}}} + \lambda = 0$. Namely~:
\[
\forall i \quad A_{i} = \frac{1}{4 \lambda^{2}}
\]

In other words~:
\[
A = \sum_{i=1}^{k} A_{i} = k \frac{1}{4 \lambda^{2}}
\]
and subsequently~:
\[
\lambda = \frac{1}{2} \sqrt{\frac{k}{A}}
\]

Assigning $\lambda$ back to $A_{i}$ yields~:
\[
\forall i \quad A_{i} = \frac{A}{k}
\]
\end{proof}

\section{The Free Perimeter $l^{\star}(A)$ for $A < \frac{1}{2}XY$}\label{sec.general}

From the proof of Theorem \ref{thm1} we saw that cutting the rectangle $D[X,Y]$ into two equal pieces by a cut parallel to the short side of length $X$ of $D[X,Y]$ is optimal w.r.t the length of the free perimeter. The results we have, in fact, state that if a shape $S$ of an area $A$ is to be separated by a short fence in $D[X,Y]$ we shall have~:
\[
\begin{array}{ccc}
 f_{FP}(S) \geq & 2 \sqrt{\pi} \sqrt{A} & \quad \emph{if $S$ touches 0 sides} \\
 f_{FP}(S) \geq & \sqrt{2 \pi} \sqrt{A} & \quad \emph{if $S$ touches 1 sides} \\
 f_{FP}(S) \geq &   \sqrt{\pi} \sqrt{A} & \quad \emph{if $S$ touches 2 adjacent sides} \\
 f_{FP}(S) \geq & 2 X                   & \quad \emph{if $S$ touches 2 opposite sides} \\
 f_{FP}(S) \geq &   X                   & \quad \emph{if $S$ touches 3 sides} \\
 f_{FP}(S) \geq & \sqrt{\pi} \sqrt{XY - A} & \quad \emph{if $S$ touches 4 sides} \\
\end{array}
\]

We shall now ask what happens when $A < \frac{1}{2}XY$, and as $A \rightarrow 0$. It is clear that for any $A$ we can separate a shape of area $A$ with a cut of size $X$, hence for every value of $A < \frac{1}{2} XY$ it holds that $l^{\star}(A) \leq X$.

Contemplating the above inequalities we realize that while $A$ is such that $\sqrt{\pi} \sqrt{A}$ is not less than $X$ we cannot hope to find a better cut! Hence, if~:
\[\sqrt{\pi} \sqrt{A} \geq X\]
namely, if~:
\[A \geq \frac{X^{2}}{\pi} \quad \emph{then}\]
we shall have~:
\[l^{\star}(A) \geq X\]

This can also be obtained using a quarter of a circle of radius $r = \frac{2X}{\pi}$.

What happens when $A < \frac{X^{2}}{\pi}$? It can be seen that from this point it pays to use quarter-circular of smaller and smaller radii, that will achieve the bound of $l^{\star}(A) = \sqrt{\pi} \sqrt{A}$. We therefore get the following result~:

\begin{theorem}  \label{thm2}
\[l^{\star}(A) = \left\{ \begin{array}{ccc}
                        X \quad & for & \  \frac{X^{2}}{\pi} \leq A \leq \frac{1}{2} XY \\
                        \sqrt{\pi A} \quad & for & \  A  \leq \frac{X^{2}}{\pi}
                      \end{array}
 \right.\]
\end{theorem}

\section{The Free Perimeter $l^{\star}(A)$ for $A > \frac{1}{2}XY$}\label{sec.general2}

Due to symmetry considerations, we can see that for any shape $S$ of area larger than $\frac{1}{2}XY$ we can simply analyze the combined free perimeters of the shapes that comprise the complement $S^{C} \triangleq D[X,Y] \setminus S = S^{C}_{1}, S^{C}_{2}, S^{C}_{3}, \ldots, S^{C}_{k}$, as it clearly equals the free perimeter of $S$. From the results shown in the previous section, we already know that the free perimeter of $S$ is minimized when $S^{C}$ is in fact a single connected shape, that touches either two adjacent sides of the rectangle, or three of its sides (depending on the area of $S$). In other words, $S^{C}$ is either a portion of the rectangle that is generated using a cut which is parallel to its shorter side, or a quarter of a circle of radius $r \leq \frac{2X}{\pi}$.

We can now complete our bound concerning the free perimeter for shapes of area larger than $\frac{1}{2}XY$, as follows~:

\begin{theorem}  \label{thm3}
\[l^{\star}(A) = \left\{ \begin{array}{ccc}
                        X \quad & for & \  \frac{1}{2} XY \leq A \leq  XY - \frac{X^{2}}{\pi} \\
                        \sqrt{\pi (XY - A)} \quad & for & \  XY - \frac{X^{2}}{\pi} \leq A  \leq XY
                      \end{array}
 \right.\]
\end{theorem}

\section{Concluding Remarks}\label{sec.conclusions}

In this paper we have completely analyzed the free perimeter isoperimetric inequality for a rectangular ambient domain. It would be very interesting to do so for various other domains as well, such as a circular domain or an annular region, and in fact any regular polygon. Our motivation for this study was a problem that arose in designing good strategies for cooperative search of smart targets using swarm of robots \cite{UAV-ROBOTICA}. As is obvious from the list of references, such problems are of great interest both from a purely geometric point of view, and in conjunction with some interesting robotics / multi agents search applications \cite{UAV-ROBOTICA,BKNS2009,DGCI}.


\begin{thebibliography}{1}

\bibitem{DGCI}
Y.~Altshuler, V.~Yanovski, D.~Vainsencher, I.A. Wagner, and A.M. Bruckstein.
\newblock On minimal perimeter polyminoes.
\newblock In {\em The 13th International Conference on Discrete Geometry for
  Computer Imagery (DGCI2006)}, pages 17--28, 2006.

\bibitem{UAV-ROBOTICA}
Y.~Altshuler, V.~Yanovsky, A.M. Bruckstein, and I.A. Wagner.
\newblock Efficient cooperative search of smart targets using uav swarms.
\newblock {\em ROBOTICA}, 26:551--557, 2008.

\bibitem{BKNS2009}
Peter Brass, Kyue~D. Kim, Hyeon-Suk Na, and Chan-Su Shin.
\newblock Escaping offline searchers and isoperimetric theorems.
\newblock {\em Comput. Geom. Theory Appl.}, 42(2):119--126, 2009.

\bibitem{CFG1991}
Hallard~T. Croft, K.J. Falconer, and Richard~K. Guy.
\newblock {\em Unsolved Problems in Geometry}.
\newblock Springer-Verlag (New York), 1991.

\bibitem{GKMS2007}
A.~Grune, R.~Klein, C.~Miori, and S.~Segura Gomis.
\newblock Chords halving the area of a planar convex set.
\newblock {\em Mathematical inequalities and applications}, 10:205--216, 2007.

\bibitem{MPS2004}
C.~Mioria, C.~Perib, and S.~Segura Gomisa.
\newblock On fencing problems.
\newblock {\em Journal of Mathematical Analysis and Applications},
  300:464--476, 2004.

\bibitem{RO1979}
Robert Osserman.
\newblock Bonnesen-style isoperimetric inequalities.
\newblock {\em The American Mathematical Monthly}, 86(1):1--29, 1979.

\end{thebibliography}
\end{document}